\documentclass[aps,superscriptaddress,amsmath,amssymb,twocolumn,amsfonts]{revtex4-1}
\usepackage{graphicx}

\usepackage{hyperref}
\usepackage{cleveref}
\usepackage{color}
\usepackage{float}
\usepackage{soul}

\usepackage{ulem}

\newcommand{\pdag}{{\phantom\dagger}}

\allowdisplaybreaks

\begin{document}

\title{Spectroscopic signatures of next-nearest-neighbor hopping in the charge and spin dynamics of doped one-dimensional antiferromagnets}
\author{Umesh Kumar}
\affiliation{Department of Physics and Astronomy, The University of Tennessee, Knoxville, Tennessee 37966, USA}
\affiliation{Joint Institute of Advanced Materials at The University of Tennessee, Knoxville, Tennessee 37996, USA}
\affiliation{Theoretical Division, T-4, Los Alamos National Laboratory, Los Alamos, New Mexico 87545, USA}

\author{Gregory Price}
\affiliation{Department of Mathematics, Augusta University, 1120 15$^{th}$ Street, Augusta, Georgia 30912, USA}
\affiliation{Department of Chemistry and Physics, Augusta University, 1120 15$^{th}$ Street, Augusta, Georgia 30912, USA}

\author{Kenneth Stiwinter}
\affiliation{Department of Chemistry and Physics, Augusta University, 1120 15$^{th}$ Street, Augusta, Georgia 30912, USA}

\author{Alberto Nocera}
\affiliation{\mbox{Stewart Blusson Quantum Matter Institute, University of British Columbia, Vancouver, British Columbia, Canada V6T 1Z4}}
\affiliation{Department of Physics and Astronomy, University of British Columbia, Vancouver, British Columbia, Canada V6T 1Z1}

\author{Steven Johnston}
\email[Corresponding author:]{sjohn145@utk.edu}
\affiliation{Department of Physics and Astronomy, The University of Tennessee, Knoxville, Tennessee 37966, USA}
\affiliation{Joint Institute of Advanced Materials at The University of Tennessee, Knoxville, Tennessee 37996, USA}

\author{Trinanjan Datta}
\email[Corresponding author:]{tdatta@augusta.edu}
\affiliation{Department of Chemistry and Physics, Augusta University, 1120 15$^{th}$ Street, Augusta, Georgia 30912, USA}

\date{\today}

\begin{abstract}
We study the impact of next-nearest-neighbor (nnn) hopping on the low-energy collective excitations of strongly correlated doped antiferromagnetic cuprate spin chains. Specifically, we use exact diagonalization and the density matrix renormalization group method to study the single-particle spectral function, the dynamical spin and charge structure factors, and the Cu $L$-edge resonant inelastic x-ray scattering (RIXS) intensity of the doped $t$-$t^\prime$-$J$ model for a set of $t^\prime$ values. 
We find evidence for the breakdown of spin-charge separation as $|t^\prime|$ increases and identify its fingerprints in the  dynamical response functions. The inclusion of nnn hopping couples the spinon and holon excitations, resulting in the formation of a spin-polaron, where a ferromagnetic spin polarization cloud dresses the doped carrier. The spin-polaron manifests itself as additional spectral weight in the dynamical correlation functions, which appear simultaneously in the spin- and charge-sensitive channels. We also demonstrate that RIXS can provide a unique view of the spin-polaron, due to its sensitivity to both the spin and charge degrees of freedom.
\end{abstract}

\maketitle
\section{Introduction}\label{Sec:Intro}
A central problem in condensed matter physics is to understand how charge and spin carriers couple to collective excitations in strongly correlated materials. For example, the behavior of a small number of holes introduced into an antiferromagnetic background of spins lays at the heart of unconventional superconductivity in the high-temperature (high-T$_c$) superconducting cuprates \cite{RevModPhys.78.17, RevModPhys.84.1383, DagottoScience1996,RevModPhys.87.457}. But it is still an open question as to how superconductivity emerges from the complex interplay of the spin, charge, and lattice excitations \cite{RevModPhys.78.17,RevModPhys.84.1383,MaierPRL2005,AndersonScience2007,PhysRevLett.100.237001,ZhengScience2017,QinarXiv,JiangarXiv,Johnston2010}. 

To address this problem, the community has developed powerful numerical approaches for simulating single- and multi-band Hubbard models and several techniques are now available for computing their ground and excited-state properties~\cite{PhysRevB.45.6479, WhitePRB1989, MaierPRL2005, MoritzNJP2009, Hanke2010, PhysRevB.92.195108, PhysRevX.5.041041, Jia2014, PhysRevX.7.031059, HuangScience2017, ZhengScience2017, QinarXiv, JiangarXiv, QinarXiv,PhysRevResearch.2.013295}. It is now possible to make detailed predictions of dynamical correlation functions in many cases, which can be compared directly with spectroscopies like angle-resolved photoemission spectroscopy (ARPES)~\cite{RevModPhys.75.473, Kim2006}, inelastic neutron scattering (INS)~\cite{MASON2001281}, and resonant inelastic x-ray scattering (RIXS)~\cite{AmentRMP2010,KotaniRMP2001}. 

Algorithmic advances have produced significant new insights into the physics of the Hubbard model itself, and a conceptual picture of competition has come into focus in recent  years~\cite{DagottoScience1995, DavisPNAS}. Here, the strong electronic correlations produce multiple nearly-degenerate states, where subtle perturbations can stabilize one state over another. For example, state-of-the-art numerical simulations of the singleband Hubbard model have found that its ground state lies very close in energy to various charge- and spin-orders (i.e. stripes) that compete with superconductivity \cite{ZhengScience2017, Jiang1424, JiangarXiv, QinarXiv, HuangScience2017}. As a result, the pure Hubbard model with only nearest-neighbor (nn) hopping $t$ does not appear to have a superconducting ground state for interaction strengths that are physically relevant for the high-T$_c$ cuprates \cite{QinarXiv}. This conclusion is extremely sensitive to perturbing factors \cite{Jiang1424, JiangarXiv, QinarXiv, HuangScience2017}, however. For example, the inclusion of next-nearest-neighbor (nnn) hopping $t^\prime$ can frustrate the stripes and stabilize $d$-wave superconductivity~\cite{JiangarXiv,QinarXiv}. It is, therefore, important to study the effects of nnn hopping and other realistic factors like disorder, lattice interactions, and additional orbitals on the properties of correlated electron models. It is also necessary to elucidate their effects on the dynamical properties of the model to be able to identify their relevance in real materials with spectroscopies. A non-zero $t^\prime$, for instance, has a measurable influence on the spin response of the 2D Hubbard model \cite{Jia2014}, which can be studied using INS or RIXS.  

Quasi-one-dimensional (1D) systems have attracted considerable attention -- both from a theoretical \cite{Ekaterina2019,QP_in_1D,PrelovekPRB1991,Eder1997,WhitePRB1998,SanoPRB2011,UKUMARNJP2018,Ekaterina2019,DagottoScience1996,UKumar2019,QinarXiv,Patel2019} and experimental \cite{BarnesPRB1993,fsNatComm,Nature.485.82, PhysRevB.88.195138,PhysRevLett.111.067204, BOUNOUA201849, PhysRevLett.118.107201, PhysRevB.95.224429, PhysRevLett.110.265502} perspective -- as these systems have traditionally been more amenable to theoretical modeling and analysis~\cite{QP_in_1D}. 
Our current understanding of the correlated 1D spin chains described by the $t$-$J$ or Hubbard model with nn hopping is built on the idea of an exotic quantum liquid that can support spinless charge (holons) and chargeless spin (spinons) excitations. Bosonization calculations supplemented with renormalization group analysis suggest that the universal fixed point for the fermionic 1D $t$-$J$ chain is the Luttinger liquid \cite{QP_in_1D}, which can support spin-charge separated (fractionalized) holon and spinon modes. But the introduction of the nnn hopping $t^{\prime}$ can spoil this clean separation of the spin and charge degrees of freedom such that it is no longer possible to write down a wave function factorized into its constituent charge and spin excitation components.

The nature of the quantum state in the 1D $t$-$J$ chain with further neighbor hopping and interactions has been studied via exact diagonalization~\cite{PrelovekPRB1991, Eder1997, SanoPRB2011} and perturbative~\cite{PhysRevB.55.15475} methods. The ground state of 
a 1D Hubbard chain with nnn hoppings was also studied with the density matrix renormalization group method (DMRG)~\cite{daul1996dmrg}. 
These studies indicate the presence of a ferromagnetic spin-polaron state in a suitable parameter regime \cite{PrelovekPRB1991, Eder1997, SanoPRB2011, PhysRevB.55.15475}. The perturbative approach points to the existence of either a ferromagnetic or antiferromagnetic state depending on the relative sign of $t$ and $t^\prime$~\cite{SanoPRB2011}. Treating the hopping of the hole to third-order in perturbation theory also suggests an effective exchange constant that scales as $J_\mathrm{eff} \sim ~ \tfrac{t(t^\prime)^2}{\epsilon^2}$. (Here, $t$ is the nn hopping, $t^\prime$ is the nnn hopping, and $\epsilon$ is the on-site energy, different from Hubbard $U$). The 1D $t$-$t^\prime$-$J$ model can be mapped onto a zig-zag chain. Hence, the bound state of the triplet state (two up spins) with a single hole can be visualized to live on the triangular plaquette \cite{Eder1997}. The physics of spin-polarons has also recently received renewed interest in the context of the 2D cuprates~\cite{PhysRevLett.106.036401}.

Early calculations of the dynamical properties of  the $t$-$t^\prime$-$J$ model and spin-polaron were carried with ED for short chains~\cite{Eder1997}. Here, we study the problem more broadly with improved momentum resolution and including the RIXS response. Our goal is to determine how nnn hopping alters the physics of a strongly-correlated antiferromagnetic spin chains and how this might be detected spectroscopically. To this end, we use density matrix renormalization group (DMRG) to compute the single-particle spectral function $A(k,\omega)$, which can be measured using ARPES, and the dynamical spin- and charge-structure factors $S(q,\omega)$ and $N(q,\omega)$, respectively, which can be measured using {\it e.g.} INS or EELS (electron energy-loss spectroscopy). RIXS also encodes information about spin- and charge- excitations, but making direct links between the RIXS intensity and $S(q,\omega)$ and $N(q,\omega)$ is not straightforward~\cite{PhysRevX.6.021020,Jia2014}. We, therefore, also explicitly compute the Cu $L$-edge RIXS response of the model using ED and the Kramers-Heisenberg formalism. In doing so, we provide predictions for the dynamical properties of the $t$-$t^\prime$-$J$ model for a range of $t^\prime/t$ values. We find evidence for the breakdown of spin-charge separation and the formation of a spin-polaron with increasing $|t^\prime|$, but which is sensitive to the sign and magnitude of the nnn hopping. Furthermore, $t^\prime$ also has an (sometimes drastic) impact on the collective excitations. We also show how the spin-polaron's presence can be identified through the appearance of additional spectral weight that forms simultaneously in the spin- and charge channels of the system's dynamical response functions. 

This paper is organized as follows: Sec.~\ref{sec:model} presents the model and methods used to study the dynamical correlation functions and RIXS cross-section. Next, Sec.~\ref{sec:ResultsandDiscussion} presents our results, beginning with a review of the non-interacting limit in Sec. \ref{sec:noninteracting}. Sec. \ref{sec:akw} focuses on the single-particle spectral function, Sec. \ref{sec:DCF} focuses on the dynamical spin- and charge-structure factors, and  Sec.~\ref{sec:CuLedge} focuses on the Cu $L$-edge spectra. Finally, Sec. \ref{sec:conclusions} provides some additional discussion and presents our conclusions.

\section{Model and Methods}\label{sec:model}
Our goal is to understand the momentum-resolved low-energy ($\le 1$~eV) collective excitations of 1D AFM cuprate spin chains. These materials have been studied successfully in the past using the $t$-$J$ Hamiltonian, where high-energy charge and orbital excitations are integrated out of a complete multi-orbital model. For example, the  $t$-$J$ model reproduces the low-energy RIXS~\cite{Nature.485.82, fsNatComm} and INS~\cite{Walters2009} spectra reported for the undoped corner-shared cuprate Sr$_2$CuO$_3$, as well as the INS data reported for the zig-zag system SrCuO$_2$~\cite{PhysRevLett.93.087202}. The use of an effective $t$-$J$ model to describe the spin-chain cuprates is also supported by a recent DMRG study that explicitly compared the RIXS spectra obtained with this model to the spectra computed from a four-orbital $pd$-model for Sr$_2$CuO$_3$~\cite{Nocera2018}. This study found that the two models agree at low-energy ($\le 1$ eV), apart from a scaling factor in their intensity that was attributed to covalency effects~\cite{Nocera2018}. 

The results mentioned above indicate that the $t$-$J$ model and its extensions can provide reliable predictions of the low-energy properties of strongly correlated cuprate spin-chains. As outlined in the introduction, however, it is also important to understand how longer-range hopping influences the results obtained from the model. To address this issue, we adopted a computational framework similar to the one used in Refs. \onlinecite{fsNatComm,UKUMARNJP2018}, but extended to include nnn hopping. The Hamiltonian is given by 
\begin{equation}\label{eq:t1t2JHamiltonian}
H=\sum_{i,j,\sigma}t^{\phantom\dagger}_{ij}\tilde{c}^\dagger_{i,\sigma}\tilde{c}^{\phantom\dagger}_{j,\sigma}+ J\sum\limits_{i}\left(\textbf{S}_i\cdot \textbf{S}_{i+1} - \frac{1}{4} n_i n_{i+1}\right).  
\end{equation}
Here, $t_{ij}=t$ and $t^\prime$ are the nn ($j=i\pm1$) and nnn ($j = i \pm 2$) hopping integrals, respectively, and $t_{ij}=0$ otherwise; $\tilde{c}^\dagger_{i,\sigma}~(\tilde{c}^{\phantom\dagger}_{i,\sigma})$ is the creation (annihilation) operator for a spin-$\sigma$ ($= \uparrow,\downarrow$) hole at site $i$ under the constraint of no double occupancy; $J$ is the exchange coupling;  $n_{i}=\sum_{\sigma}\tilde{c}^\dagger_{i,\sigma}\tilde{c}^{\phantom\dagger}_{j,\sigma}$ is the number operator; and $\textbf{S}_{i}$ is the spin operator at site $i$.  To facilitate comparisons with previous work, we adopt model parameters $t=0.4$ eV, $J=0.25$ eV, which are typical for corner-shared cuprate materials~\cite{UKUMARNJP2018}, and vary $t^\prime$ between $[-\tfrac{t}{2}, \tfrac{t}{2}]$. 

We access the model's collective excitations by computing several dynamical correlation functions. The first is the single-particle spectral function 
\begin{equation}\label{eq:Akw}
A(k,\omega) = \sum_{f,\sigma}\left| \langle f | \tilde{c}^\dagger_{k,\sigma}|g\rangle \right|^2 \delta(E_f - E_g - \omega), 
\end{equation}
where $\tilde{c}^\dagger_{k,\sigma}=\frac{1}{\sqrt{N}}\sum_i e^{-ikR_i}\tilde{c}^\dagger_{i,\sigma}$ is the Fourier transform of the creation operator, and $R_i$ is the lattice vector for the magnetic atom in unit cell $i$, which is associated with Cu in our case. We also considered the two-particle dynamical spin and charge structure factors, which are 
defined as
\begin{equation}\label{Eq:Sqw}
S(q, \omega) =  \sum_{f}\left\vert\langle f\vert \hat{S}^\alpha_q\vert g\rangle\right\vert^2 \delta (E_f-E_g-\omega)
\end{equation}
and
\begin{equation}\label{eq:Nqw}
N(q, \omega) = \sum_{f}\left\vert\langle f\vert 
\hat{N}_q\vert g\rangle\right\vert^2 \delta (E_f-E_g-\omega),
\end{equation}
respectively. Here, $q$ and $\omega$ are the 1D momentum and energy transfer to the chains, respectively, and $\hat{S}^\alpha_q = \frac{1}{\sqrt{N}}\sum_{i=1}^{N} e^{-iqR_i}S_i^\alpha$ and $\hat{N}_q=\frac{1}{\sqrt{N}}\sum_{i=1}^{N} e^{-iqR_i}n_i$ are the Fourier transforms of the $S_i^\alpha$ local spin ($\alpha=z,\pm$) and  $n_i=\sum_{\sigma}\tilde{c}^\dagger_{i,\sigma}\tilde{c}^{\phantom\dagger}_{i,\sigma}$ is the number operator, respectively. 

The RIXS intensity $I(q,\omega)$ is computed using the Kramers-Heisenberg formalism, and is given by
\begin{equation}
\label{eq:rixs}
I(q,\omega) \propto \sum\limits_{f}\left\lvert
M_{fg}
\right\rvert^{2}\delta\left(E_{f} - E_{g} - \omega\right),
\end{equation}
where
\begin{equation*}
M_{fg} = 
\sum\limits_{n}\frac{\langle f \lvert D^{\dagger}_{k_\mathrm{out}}\rvert n \rangle
\langle n \lvert D^{\phantom\dagger}_{k_\mathrm{in}} \rvert i \rangle}
{E_{g} + \omega_\mathrm{in} - E_{n} + i \Gamma_{n}}.
\end{equation*}
In the above expression,
the incoming (outgoing) photons  have energy $\omega_\mathrm{in}$ ($\omega_\mathrm{out}$) and momentum $k_\mathrm{in}$ ($k_\mathrm{out}$); 
$\omega = \omega_\mathrm{in} - \omega_\mathrm{out}$ and 
$q = k_\mathrm{in}-k_\mathrm{out}$ are the energy and momentum transferred along the chain direction, respectively; $\lvert g
\rangle$, $\lvert n \rangle$, and $\lvert f \rangle$ are the initial, intermediate, and final states of the RIXS process with energies $E_{g}$, $E_{n}$, and $E_{f}$, respectively; $D_{\bf k}$ is the dipole operator describing the $2p\rightarrow 3d$ atomic transition; and  $\Gamma_n$ is related to the inverse core-hole lifetime. For our numerical calculations we used $\Gamma_n = 0.3$ eV, independent of the value of $n$, as is appropriate for the Cu $L$-edge~\cite{Jia2014}.

At the Cu $L$-edge, the dipole operator takes the form  
\begin{equation}\label{Eq:Dipole_CuL}
D_{k} = \sum\limits_{i\sigma}e^{ik\cdot{R}_{i}}\left[\tilde{d}^\pdag_{i,\sigma}p^{\dagger}_{i,\sigma} + \mathrm{h.c.} \right],
\end{equation} 
where $p^\dagger_{i,\sigma}$ ($p^{\phantom\dagger}_{i,\sigma}$) creates (annihilates) a spin $\sigma$ hole in a Cu $2p$ orbital located at site $i$. Note that we have neglected an orbital-dependent prefactor in Eq. \eqref{Eq:Dipole_CuL} that depends on the photon polarization and scattering geometry. In what follows, we will consider the spin-conserving and non-spin-conserving channels  individually. We, therefore, also take into account the effect of the spin-orbit coupling in the description of the $2p$ core states.

For all of the RIXS spectra shown in this work, we set the incident photon  $\omega_\mathrm{in}$ to coincide with the resonance (maximum intensity) of the XAS obtained for the same model. Here, $I_\mathrm{XAS}(\omega)$ is computed using Fermi's golden rule and is given by 
\begin{equation}\label{eq:XAS}
I_\mathrm{XAS}(\omega)\propto\sum\limits_{n}\left\lvert\langle n \lvert D_{k = 0} \rvert g \rangle\right\rvert^{2}
\delta\left(E_{n} - E_{g} - \omega\right).
\end{equation}

The single particle spectral function and dynamical spin and charge structure functions are computed using the DMRG correction-vector method \cite{PhysRevB.60.335} and the Krylov decomposition \cite{PhysRevE.94.053308}, as implemented in the DMRG++ code \cite{alvarez0209}. This approach requires real-space representations of Eqs. \eqref{eq:Akw} -- \eqref{eq:Nqw}, which can be found in Ref. \onlinecite{Nocera2018b}. Our DMRG calculations were carried out on $N = 80$ site chains with open
boundary conditions with a fixed number of holes ($N_h=76$) such that $\langle n \rangle = \frac{N_h}{N}=0.95$. We also kept up to $m = 1000$ DMRG states to maintain a truncation error below $10^{-7}$ and introduced a spectral broadening in the correction-vector approach fixed at $\eta = 0.08t$. To compute the XAS and RIXS intensities, we diagonalized Eq.~\eqref{eq:t1t2JHamiltonian} exactly on $N=20$ site chains with periodic boundary conditions and used the eigenstates to evaluate Eqs. \eqref{eq:rixs} and \eqref{eq:XAS}. When numerically evaluating Eq. \eqref{eq:rixs}, we approximated the energy-conserving delta function with a Gaussian function $\delta(\omega) \approx \frac{1}{\gamma\sqrt{2\pi}}e^{-\frac{\omega^2}{2\gamma^2} }$, with $\gamma = \tfrac{t}{10}$. 

\begin{figure}[t]
    \includegraphics[width=\columnwidth]{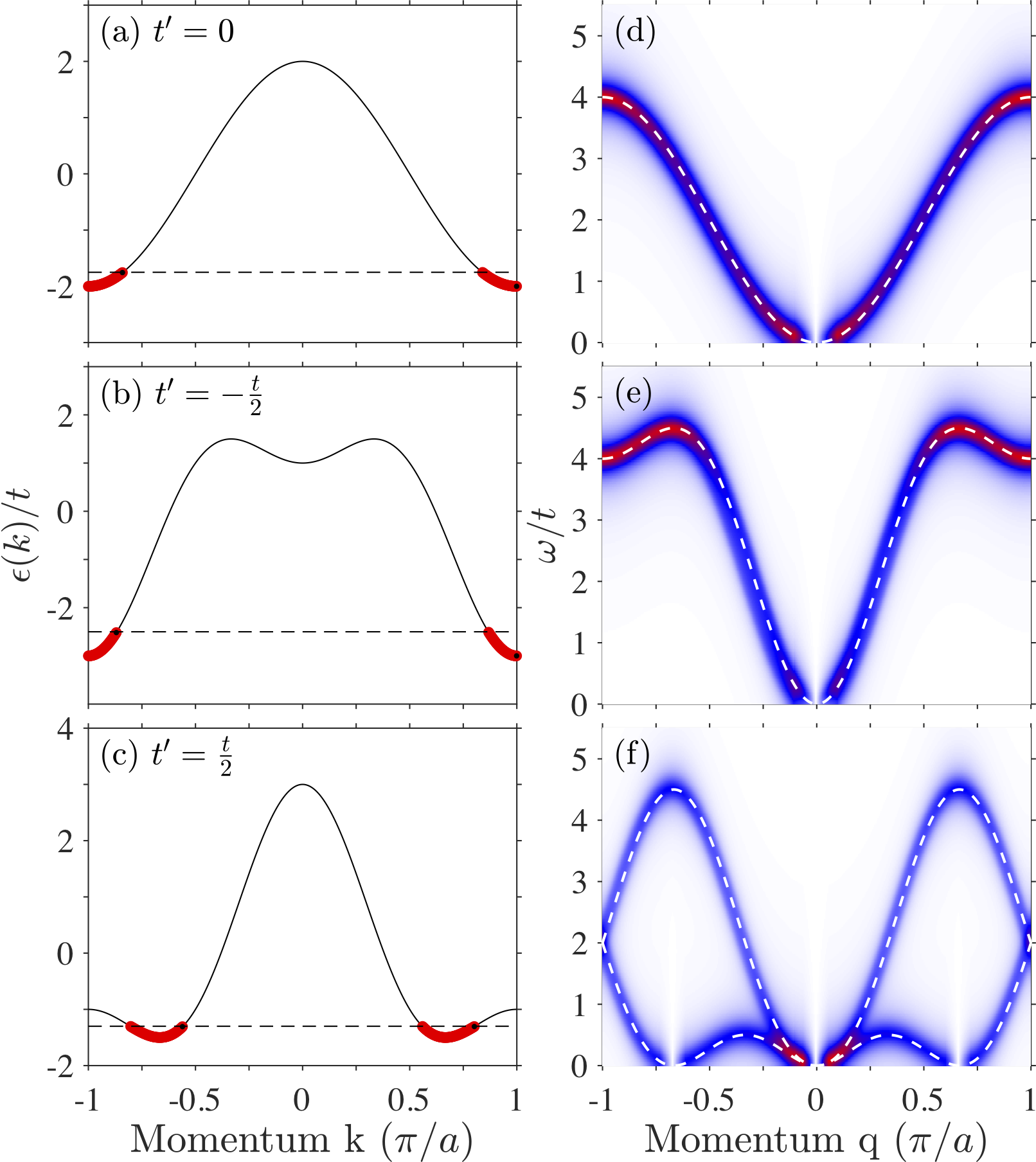}
    \caption{Results for the noninteracting 1D $t$-$t^\prime$ model, calculated on an $N=400$ site chain. Panels (a)~-~(c) show the band dispersion of the non-interacting model $\epsilon(k)=2t\cos(ka)+2t^\prime\cos(2ka)$ for $t^\prime=0$,$-\tfrac{t}{2}$, and $\tfrac{t}{2}$, respectively. The thick red overlays indicate the location of the filled states in the dilute limit and the dashed line indicates the resulting Fermi energy. Panels (d)~-~(f) show the corresponding dynamical charge structure factors $N(q,\omega)$ calculated on an $N = 400$ site cluster and a total filling of $\langle n \rangle = 0.05$. The overlays in panels (d)~-~(f) are given by $\omega(q)=\epsilon(k)-\epsilon(k_\text{min})$ and plotted as a function of $q = k-k_\mathrm{min}$.}
    \label{fig:noninteractingband}
\end{figure}

\section{Results and Discussion\label{sec:ResultsandDiscussion}}
\subsection{Results in the non-interacting limit}\label{sec:noninteracting}
Before proceeding to our main results, it is useful to examine the single- and two-particle responses for the non-interacting model. Our aim here is to remind the reader of how $t^\prime \ne 0$ alters the bare band structure and the topology of the Fermi surface~\cite{PhysRevB.96.085133}. 

Figure \ref{fig:noninteractingband} shows results for the single-particle dispersion and dynamical charge structure factor $N(q,\omega)$. The excitation spectrum in this case is determined by the dispersion $\epsilon(k) = 2t\cos\left(ka\right)+ 2t^\prime \cos\left(2ka\right)$, which is plotted as thin black lines in Figs.~\ref{fig:noninteractingband}(a)-(c) for $t^\prime/t = 0$, $-\tfrac{1}{2}$, and $\tfrac{1}{2}$, respectively. When $t^\prime = 0$, the band structure has a simple cosine shape, with a local minimum located at the zone boundary, consistent with the use of hole language in Eq.~\eqref{eq:t1t2JHamiltonian}. The band structure is modified for $t^\prime \ne 0$ (throughout, we assume that $|t^\prime| \le |t/2|$). For example, the band maxima shift from zone center to $k_\mathrm{min}=\pm\tfrac{1}{a}\cos^{-1}(\tfrac{t}{4t^\prime})$ when $t^\prime < -\tfrac{t}{4}$ [Fig. \ref{fig:noninteractingband}(b)], while the local minima shifts from the zone boundaries to $k_\mathrm{min}=\pm\tfrac{1}{a}\cos^{-1}(-\tfrac{t}{4t^\prime})$ when $t^\prime > \frac{t}{4}$ [Fig. \ref{fig:noninteractingband}(c)]. These changes in the locations of the band extrema alter the topology of the Fermi surface when the band is nearly empty or nearly filled. This fact is illustrated by the thick red lines in Fig.~\ref{fig:noninteractingband}, which indicate the occupied states in the case of a nearly empty band. One can see that the Fermi surface transitions from a single pocket centered at the zone boundary for $t^\prime \le \tfrac{t}{4}$ to two Fermi surfaces centered at $k_\mathrm{min}\ne\pm \tfrac{\pi}{a}$ when $t^\prime > \tfrac{t}{4}$. 

The topological change of the Fermi surface can have a profound effect on the excitation spectrum of the system. To illustrate this, Figs.~\ref{fig:noninteractingband}(d)-(f) plots $N(q,\omega)$ obtained when $\langle n \rangle = 0.05$ spinless fermions/unit cell occupy the bands shown in Figs.~\ref{fig:noninteractingband}(a)-(c), respectively. (We consider spinless fermions to make a connection to holon excitations in the doped interacting chains, see below and also App.~B of Ref.~\cite{UKUMARNJP2018}.) The overlays in Figs. ~\ref{fig:noninteractingband}(d)-(f) are guides to the eye; they are given by $\omega(k-k_\text{min}) = \epsilon(k) - \epsilon(k_\text{min})$, where $k_\text{min}$ is the momentum of the lowest filled state in each case. In other words, the overlay represents scattering from the band minimum to another $k$ value. In this case, $N(q,\omega)$ measures intraband scattering and provides indirect information on the underlying band structure. 
The charge response in panels (d) and (e) consist of a single dispersing narrow continuum. In contrast, panel (f) has two distinct branches, which occurs because scattering to the left and the right from each of the Fermi surfaces is no longer equivalent when $t^\prime = \tfrac{t}{2}$ and the band minima shift away from the high-symmetry points. Many of these features persist in the interacting system (see below), but the electronic interactions alter the underlying band dispersions. 

\subsection{The single-particle spectral function for the interacting case}\label{sec:akw}
We now turn our attention to the electronic properties and spectral functions of the interacting chains. The physical behavior of the 1D $t$-$J$ model without next-nearest neighbor hopping is well established~\cite{PhysRev.128.2131,PhysRevB.55.15475,PhysRevB.55.12510, 1742-5468-2006-12-P12013}. Its elementary excitations are collective spin and charge density excitations that can be mapped onto fractionalized quasiparticle excitations. The spin density excitations map to chargeless spin-$\tfrac{1}{2}$ quasiparticles known as spinons with a dispersion relation given by $\omega_\text{s}(k) = \tfrac{\pi J}{2}\sin(k a)$. Similarly, the charge density excitations map to spinless charge-$e$ quasiparticles known as holons with a dispersion $\omega_\text{h}(k) = 2t\left[1-\cos(ka)\right]$. 

\begin{figure}[t]
    \includegraphics[width=\columnwidth,angle =0]{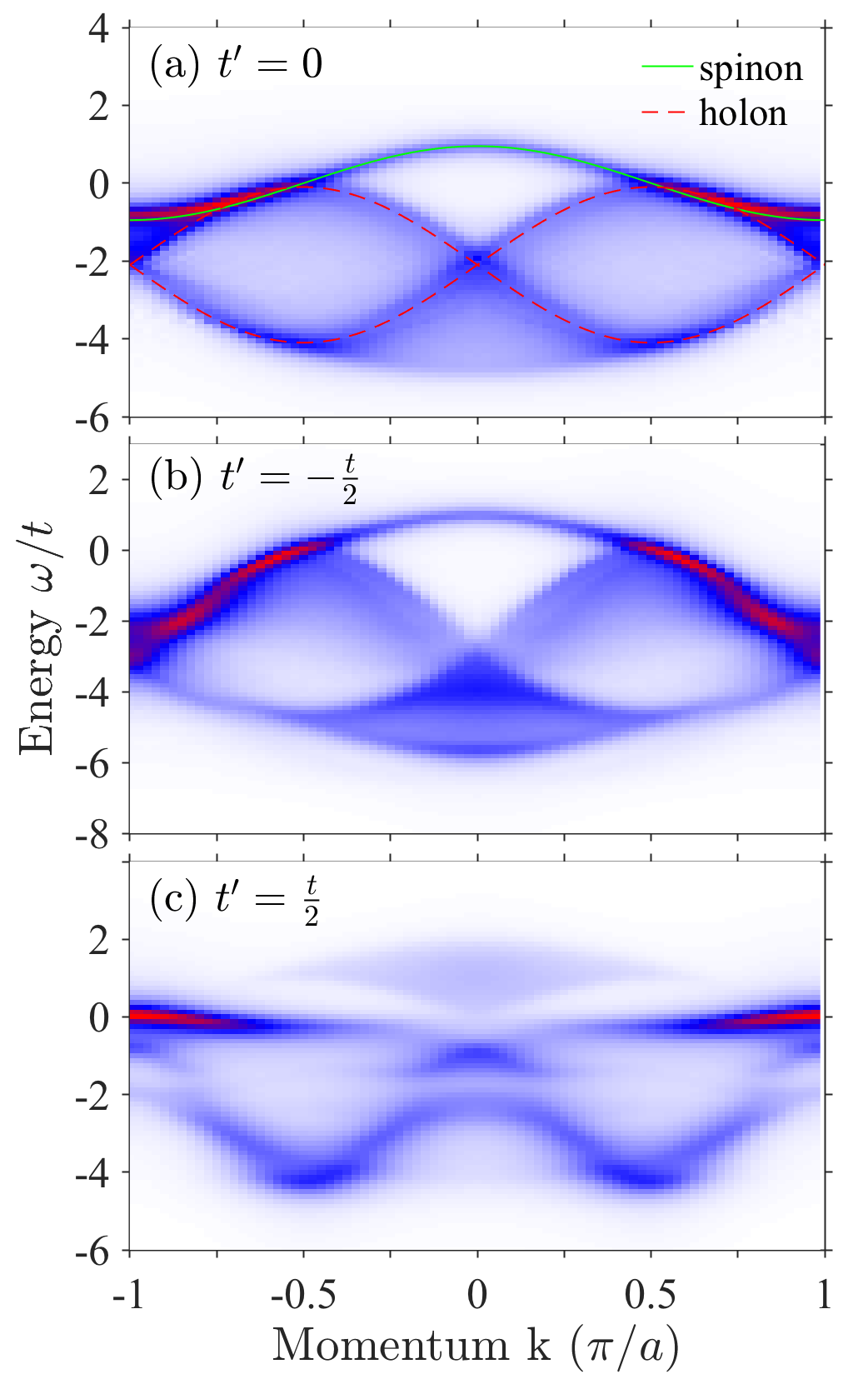}
    \caption{DMRG results for the single particle spectral function $A(k,\omega)$ of the $t$-$t^\prime$-$J$ model. The spectra were calculated on an $N = 80$ site chain with $t=0.4$~eV, $J = 0.25$ eV, $\langle n \rangle = 0.95$, and values of $t^\prime$ as indicated in each panel. The solid green and dashed red lines in panel (a) are guides for the eye for the spinon and holon excitations, respectively.}
    \label{fig:Akw}
\end{figure}

Figure \ref{fig:Akw} shows our DMRG results for the single-particle spectral function $A(k,\omega)$ for $\langle n \rangle = 0.95$ and  values of $t^\prime$, as indicated. When $t^\prime = 0$, $A(k,\omega)$ has two distinct dispersing features, which correspond to the expected spinon and holon excitations. The dispersions of these quasiparticles are highlighted using the solid green and dashed red lines, which trace the expected path. These spectral features are a fingerprint of spin-charge separation and have been directly observed in SrCuO$_2$ and other 1D spin chains using ARPES~\cite{PhysRevLett.77.4054,Kim2006,PhysRevB.59.7358,PhysRevB.73.201101,PhysRevB.56.15589}. Moreover, the spectrum resembles the main 
features of the spectral function of the 1D Hubbard model at $U/t \gg 1$~\cite{Benthien2004,Nocera2018c}.
In particular, we can attribute the spectral weight above the Fermi level to the spinon-antiholon continuum of empty states due to the anomalous spectral weight transfer upon hole-doping (electron doping in this work since we are assuming hole language).

The spectral function for $t^\prime =-\tfrac{t}{2}$ is qualitatively similar in that two distinct sets of excitations are observed that resemble the spinon and holon excitations in the $t^\prime = 0$ case.  Their dispersions, however, are modified from the pure cosine and sine forms, and a slight kink-like feature appears in the dressed dispersion relationship where the spinon and holon excitations overlap. We attribute these changes to a coupling between the spin and charge degrees of freedom introduced by the non-zero $t^\prime$ [see Fig. \ref{fig:noninteractingband}(b)]. We will show later that the changes in the quasiparticle dispersions can also be observed in the two-particle correlation functions. 

The situation is much different when $t^\prime = \tfrac{t}{2}$, as shown in Fig. \ref{fig:Akw}(c). In this case, the spectral weight is completely reorganized, and one can no longer make out spectral features for the spinons and holons. Instead, we find two new distinct sets of features. The first is a sharp peak located near $E_\mathrm{F}$ and centered at the zone boundary. This excitation is quasi-particle-like and has a very narrow bandwidth, whose spectral weight is suppressed as $k \rightarrow 0$. The second feature is a more incoherent but still quasi-particle-like excitation located between $\tfrac{\omega}{t} \in [-4,0]$. This excitation has a cosine-like dispersion with a period of $\tfrac{\pi}{a}$. 
These changes are a clear indicator that the spin-charge separation picture has broken down for this value of $t^\prime$. In the next section, we will show that the dynamical spin and charge structure factors support this conclusion. Our results agree with several previous studies~\cite{Eder1997,PrelovekPRB1991,SanoPRB2011} that found evidence for binding of the spinons and holons using similar models and parameters. 

\begin{figure}[t]
    \includegraphics[width=\columnwidth,angle =0]{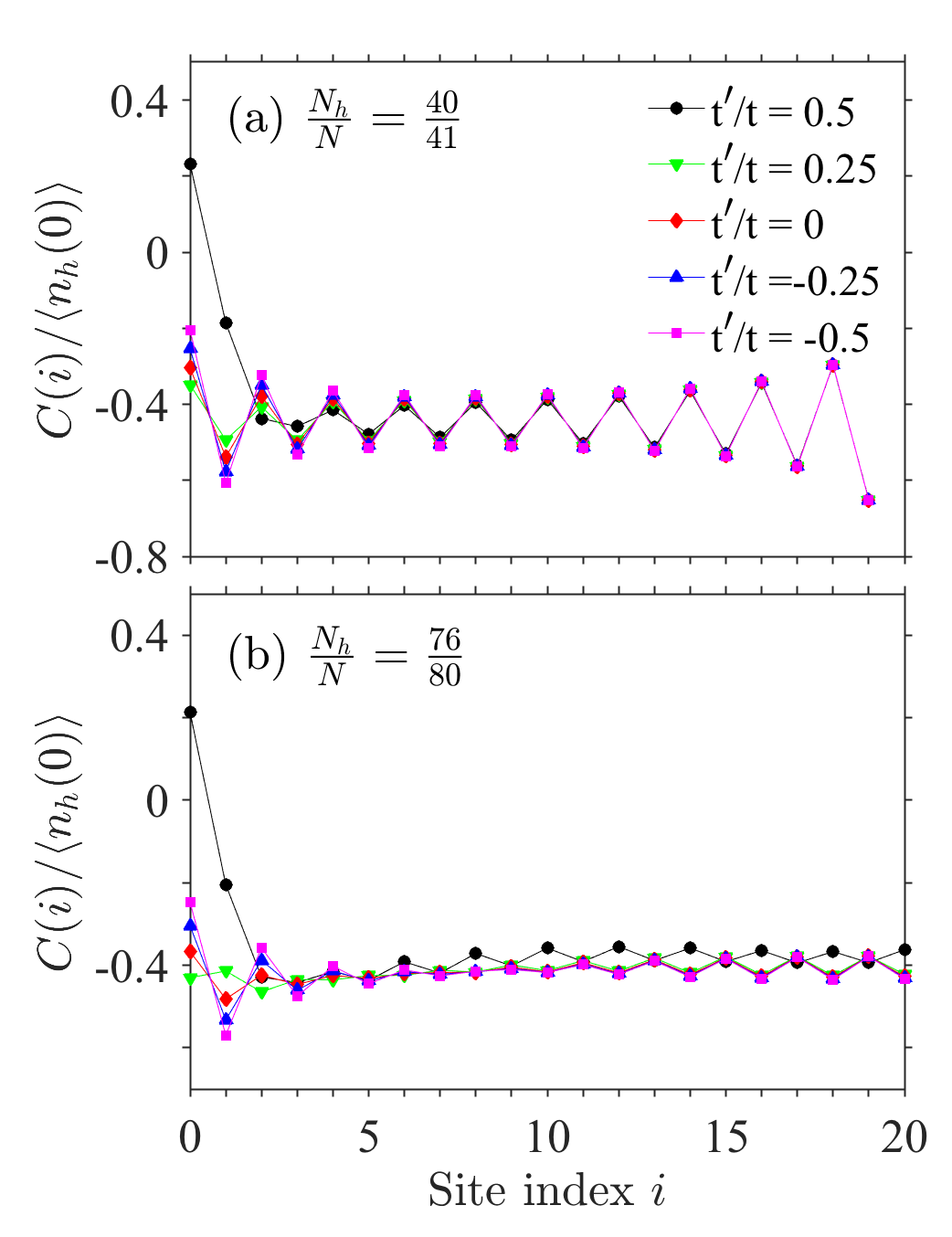}
    \caption{DMRG results for the spatial dependence of the spin-polaron correlation function for $t = 0.4$, $J = 0.25$, and as a function of $t^\prime$. Results are shown for (a) $N_h={40}$ on a $N=41$ site chain and 
    (b) $N_h={76}$ on a $N=80$ site chain.}
    \label{fig:Ci}
\end{figure}

The introduction of $t^{\prime}$ can be viewed as converting the lattice geometry from a chain to a triangular ladder, thus creating an intermediate structure between 1D and 2D. In such a system, the physical properties are known to deviate significantly from the spin-charge separation scenario, which is strictly valid in 1D only ($t^\prime=0$).
From this point of view, it is interesting to compare our results with those obtained in 2D. For example, our results in  Fig. \ref{fig:Akw}c resemble the main features of the $U/t \gg 1$ spectral function of the 2D Hubbard model~\cite{Preuss1995,Grober2000,Kohno2012,MoritzNJP2009}, where a prominent quasiparticle band appears due to the anomalous
spectral weight transfer from the upper Hubbard band to the lower Hubbard band upon hole doping~\cite{Eskes1991}.  
A similar feature was also identified as a {\it flat band} in early studies in the 2D $t$-$J$ model (with $t^\prime=0$) upon hole doping~\cite{Poilblanc1993,Dagotto1994}. 
In our case, we attribute the narrow band around the Fermi level ($\omega = 0$) with quasiparticle excitations with weakly ferromagnetic character (as shown below in our ground state analysis for ferromagnetic-polaron correlation functions in Fig.\ref{fig:Ci}b). Fluctuations arising from the Nagaoka ferromagnetism~\cite{Nagaoka1966} could dominate the antiferromagnetic fluctuations near the Mott transition upon doping, 
in the extremely large $U/t\rightarrow\infty$ limit of the Hubbard model. As we will see below when discussing the spin excitation spectrum, this feature appears as a faint excitation below the two-spinon continuum for small momentum transfer $q\simeq 0$. In the charge channel, it appears as a sharp low energy mode with spectral weight concentrated at zone center, with a width of the order of the exchange interaction $J$. In summary, our results for $t^\prime=t/2$ deviate substantially from strictly spin-charge separated 1D character, indicating that the doped charge is no longer fractionalizing, or that $t^\prime \ne 0$ induces an effect similar to a dimensional crossover towards 2D.

As previously mentioned, the introduction of a non-zero $t^\prime$ is expected to produce a  spin-polaron~\cite{Eder1997,PrelovekPRB1991,SanoPRB2011}. Here, the doped electron is dressed by a spin polarization cloud whose spatial extent depends on the strength of the exchange coupling. Specifically, the size of the ferromagnetic polarization cloud around a single hole grows as $J/t$ decreases, eventually extending across the entire system when $J/t \rightarrow 0$ \cite{PrelovekPRB1991}. To confirm the presence of the spin-polaron in our case, and to determine its spatial extent, we computed the spin-polaron correlation function 
\begin{align}\label{eq:spinpolaron}
C(i)  = \left\{ \begin{array}{ll}
\langle {\bf S}_{c-1}\cdot  n^{h}_c {\bf S}_{c+1}\rangle / \langle n^{h}_{c}\rangle & \mbox{if } i = 0\\
\langle n^{h}_c {\bf S}_i\cdot {\bf S}_{i+1}\rangle / \langle n^{h}_{c}\rangle   & \mbox{if } i > 0 
\end{array} \right.
\end{align}
using DMRG. Here, $n^{h}_{c} = 1 - n_{c}$ measures the {\it electron} occupation at the reference site index $c$ (taken to be the 
center site of the chain, unless otherwise stated), and $i$ is a site index measured relative to $c$. 
Notice that by definition, $C(i=0)$ indicates the spin correlation {\it across} the doped electron in the chain\cite{Scalapino1997,Patel2017}.
Fig. \ref{fig:Ci}(a) shows results for a single doped {\it electron} added to an $N = 41$ site chain. When $t^\prime < \tfrac{t}{2}$, $C(i)$ is negative at all distances, indicating that the spins form an antiferromagnetic background that is largely independent of the hole's position. However, $C(i)$ increases slightly within two unit cells of the doped electron and when $t^\prime = \tfrac{t}{2}$, $C(i)$ even changes sign at these distances. This behavior reflects the formation of a small ferromagnetic polarization cloud surrounding the charge. It also confirms that the spinons and holons become weakly coupled for $t^\prime < \tfrac{t}{2}$ but more strongly coupled when $t^\prime = \tfrac{t}{2}$, leading to the formation of a spin-polaron. These differences also explain why we see a more drastic reorganization of the spectral weight in Fig. \ref{fig:Akw} when $t^\prime = \tfrac{t}{2}$. 
Fig. \ref{fig:Ci}(b) shows analogous results for an $N = 80$ site chain at 5\% doping. The similarity between the two panels shows that the spin-polaron picture persists at this doping level.    

\subsection{The dynamical spin and charge structure factors}\label{sec:DCF}
We now consider the effects of $t^\prime$ on the dynamical spin $S(q,\omega)$ and charge $N(q,\omega)$ structure factors. These two-particle correlation functions can provide crucial information about the nature of the excitations in the model. We can also compare and contrast their behavior with a more complicated RIXS response. 

\begin{figure}[t]
    \includegraphics[width=\columnwidth,angle =0]{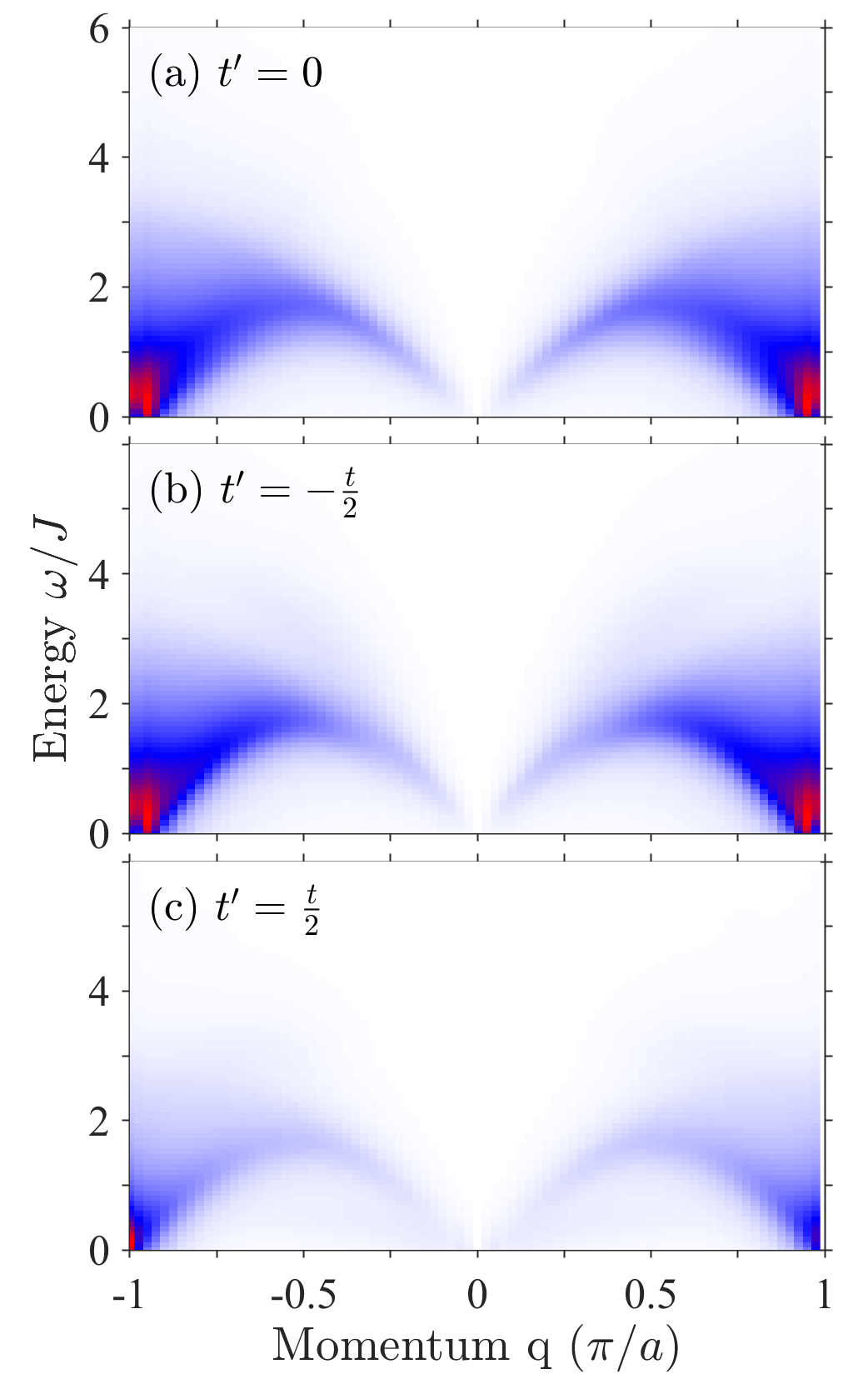}
    \caption{DMRG results for the dynamical spin structure factor $S(q,\omega)$ for the $t$-$t^\prime$-$J$ model. The spectra were calculated on an $N = 80$ site chain with $t=0.4$~eV, $J = 0.25$ eV, $\langle n \rangle = 0.95$, and values of $t^\prime$ as indicated in each panel.}
    \label{fig:Sqwt1t2model}
\end{figure}

\begin{figure}[t]
    \includegraphics[width=\columnwidth,angle =0]{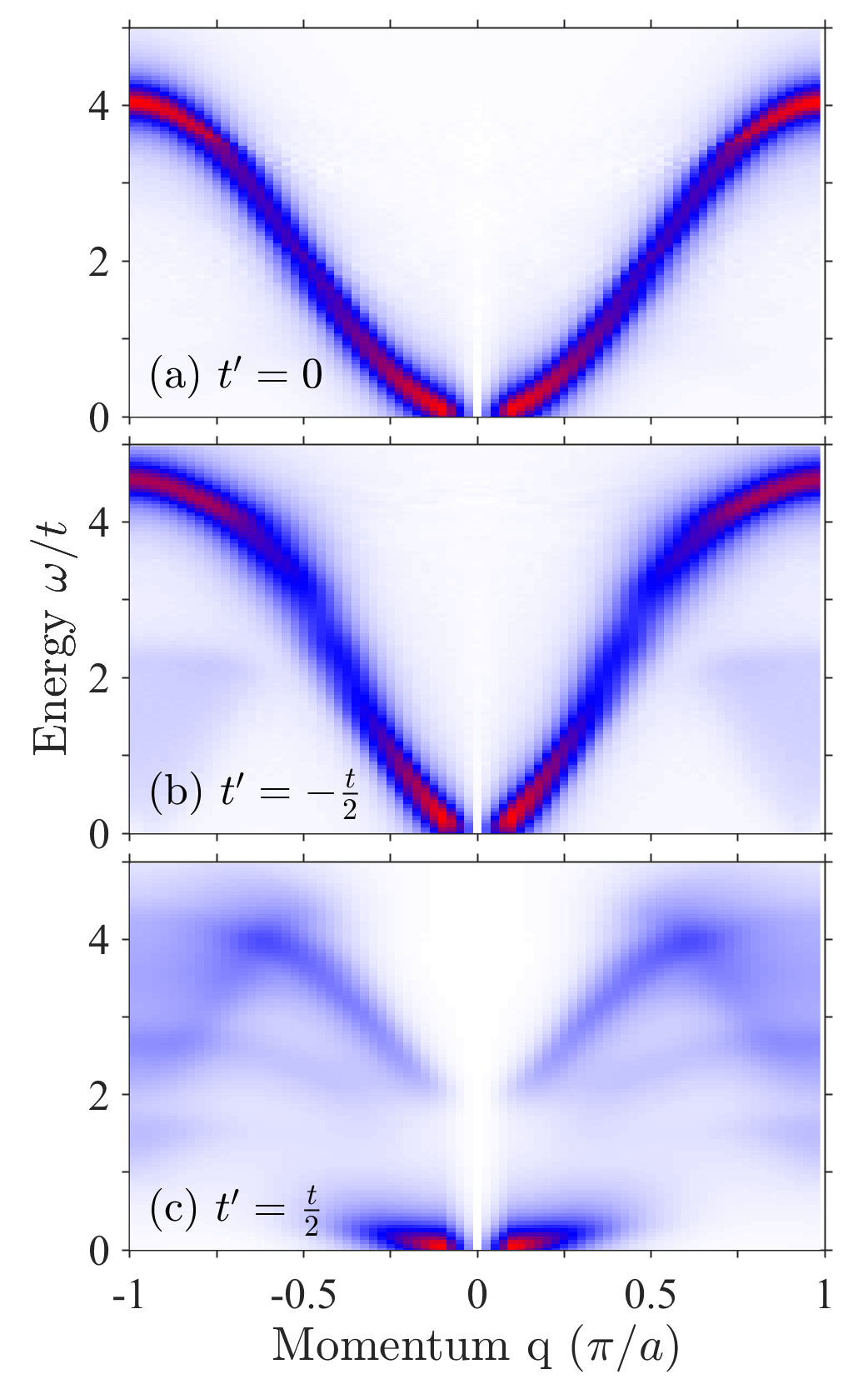}
        \caption{DMRG results for the dynamical charge structure factor $N(q,\omega)$ for the $t$-$t^\prime$-$J$ model. The spectra were calculated on an $N = 80$ site chain with $t=0.4$~eV, $J = 0.25$ eV, $\langle n \rangle = 0.95$, and values of $t^\prime$ as indicated in each panel.}
    \label{fig:Nqwt1t2model}
\end{figure}

Figure~\ref{fig:Sqwt1t2model} shows DMRG results for $S(q,\omega)$ at $\langle n \rangle = 0.95$ with the same parameters used to produce Fig. \ref{fig:Akw}. Panel (a) shows results for $t^\prime = 0$, which exhibits the usual two-spinon continuum~\cite{PhysRevB.55.12510, FOUR, Nature.485.82, Walters2009}. We find that the gapless excitation near $q=0$ broaden somewhat relative to the undoped case. At the same time, the gapless excitations near $q = \pm \tfrac{\pi}{a}$ in the undoped case shift to $\pm(\tfrac{\pi}{a}-2k_\mathrm{F})$, where 
\{$k_\mathrm{F} = \tfrac{\pi}{2a}\left(1-\langle n \rangle\right)$ is the Fermi momentum measured relative to the band minimum, 
as expected for a Luttinger liquid. These results are fully consistent with prior results for the doped $t$-$J$ and Hubbard models \cite{PhysRevB.96.195106,Ekaterina2019}. 

Figures ~\ref{fig:Sqwt1t2model}(b) and \ref{fig:Sqwt1t2model}(c) show results for the same system but now with $t^\prime=-\tfrac{t}{2}$ and $\tfrac{t}{2}$, respectively. 
In both cases, the spin excitations manifest as a continuum, but appear to harden (soften) for $t^\prime =-\tfrac{t}{2}$ ($\tfrac{t}{2}$) relative to the $t^\prime=0$ case, also in agreement with prior studies~\cite{Eder1997,PhysRevB.77.115102, Ekaterina2019}. However, we also find that the gapless excitations located at $\pm(\tfrac{\pi}{a}-2k_\mathrm{F})$ for $t^\prime = 0$ shift back to $\pm \tfrac{\pi}{a}$ when $t^\prime = \tfrac{t}{2}$, indicating that the system is deviating from the expectations for a Luttinger liquid. We also observe additional dispersing features with weak intensity outside of the boundaries of the two-spinon continuum when $t^\prime \ne 0$. Specifically, for $t^\prime = -\tfrac{1}{2}$  ($\tfrac{1}{2}$), there is a faint excitation located just above (below) the spinon continuum. Both sets of excitations disperse toward zero energy as $q \rightarrow 0$ and have corresponding features in the dynamical charge structure factor. We will return to these features in a moment. 

\begin{figure*}[t]
    \centering
    \includegraphics[width=\textwidth]{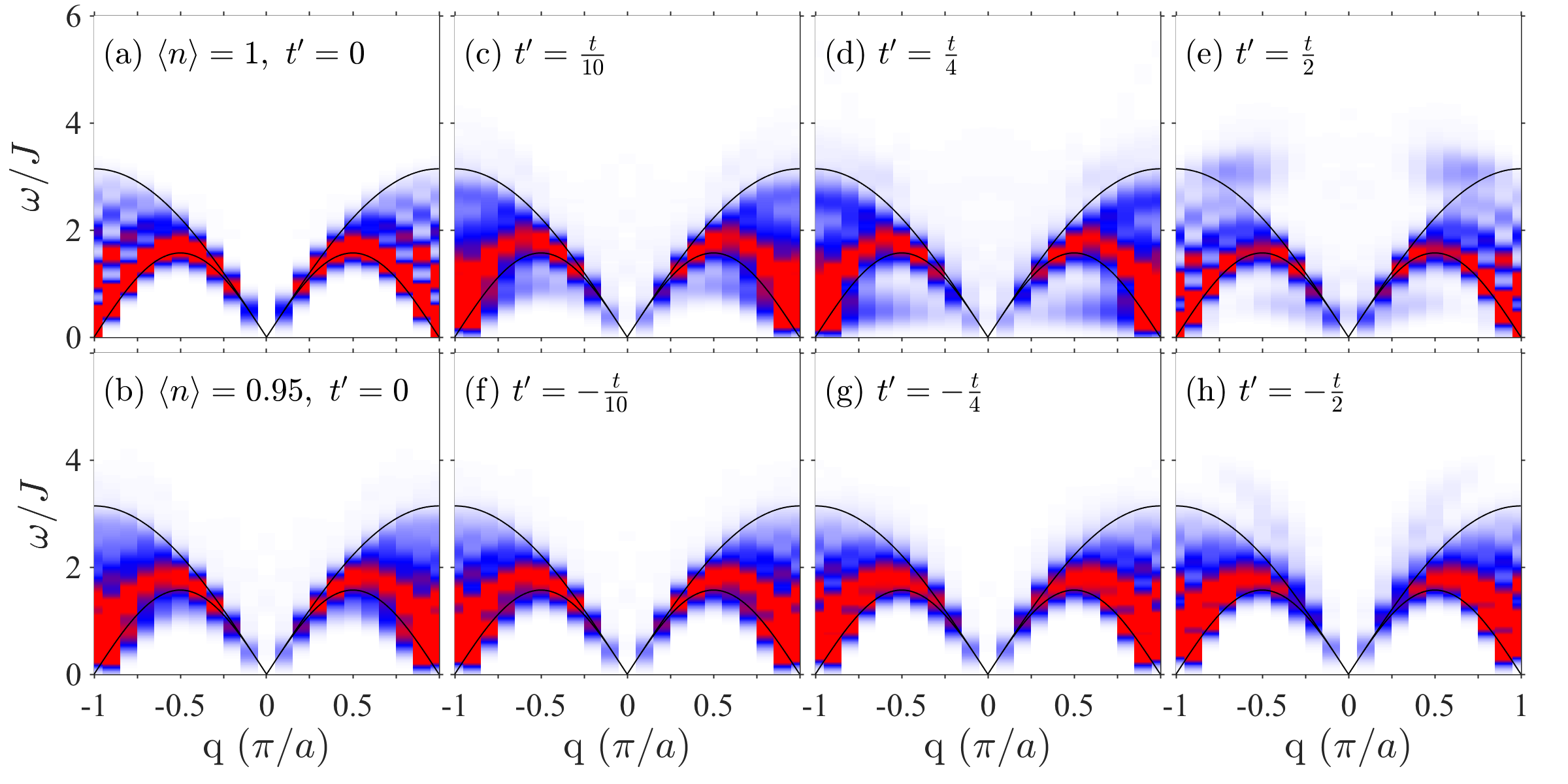}
    \caption{Exact diagonalization results for the Cu $L$-edge RIXS spectra of the the $t$-$t^\prime$-$J$ model 
    in the non-spin-conserving  ($\Delta S =1$), calculated on an $N  = 20$ site chain with $t = 0.4$ eV and $J = 0.25$ eV. Panel (a) shows results for  half-filling $\langle n \rangle = 1$ with $t^\prime = 0$, whereas panels (b)-(h) show results for $\langle n \rangle = 0.95$, and $t^\prime$ as indicated. The solid black lines in each panel indicate the boundaries of two-spinon continuum expected at half-filling when $t^\prime = 0$. For $t^\prime/t>0$ ($<0$), the continuum of spin excitations appear to soften (harden) with increasing $|t^\prime|$. At the same time, new weak excitations appear outside of the spinon continuum when $|t^\prime|$ becomes large. All panels in this figure have been plotted with the same color scale.}
    \label{fig:rixst1t2s1}
\end{figure*}
Figure~\ref{fig:Nqwt1t2model} shows the dynamical charge structure factor $N(q, \omega)$, again for $\langle n \rangle = 0.95$ and $t^\prime$ as indicated in each panel. For $t^\prime = 0$ [Fig. \ref{fig:Nqwt1t2model}(a)], the charge excitations agree well with the $N(q,\omega)$ obtained from the spinless model with $\langle n \rangle \approx 0.05$ [Fig. \ref{fig:noninteractingband}(d)]. This result is to be expected if the charge quantum number of the small number of doped carriers are carried by spinless holons. 

The situation appears to be qualitatively similar when $t^\prime =-\tfrac{t}{2}$ 
[Fig. \ref{fig:Nqwt1t2model}(b)] in that $N(q,\omega)$ has a sharp dispersing feature that loosely resembles the spectrum in \ref{fig:noninteractingband}(e). A closer inspection, however, reveals two key 
differences: First, the precise dispersion relation in Fig.~\ref{fig:Nqwt1t2model}(b) 
differs from the non-interacting case [\ref{fig:noninteractingband}(e)]. Specifically, the local maximum  remains at the zone boundaries and the overall dispersion in $N(q,\omega)$ has a slight kink-like bend near $(q,\omega) = \left(\tfrac{\pi}{2a}, 3t\right)$ in the interacting case. Interestingly, these features can also be made out in the renormalized holon dispersion shown in Fig. \ref{fig:Akw}(b), indicating that the sharp mode in $N(q,\omega)$ can be viewed as a renormalized holon excitation.   
The second significant difference is an additional weak continuum of spectral weight located below $\omega \le 2t$ and near the zone boundaries, which overlaps with the continuum of spin excitations seen in $S(q,\omega)$. The phase space overlap of these features in the two correlation functions reflects the fact that the spin and charge degrees of freedom are coupled to some extent, consistent with the formation of a weakly dressed spin-polaron. 

The charge excitations are more significantly reorganized when $t^\prime = \tfrac{t}{2}$, as shown in Fig. \ref{fig:Nqwt1t2model}(c). In this case, we observe two distinct sets of excitations, as well as a large amount of incoherent spectral weight at higher energy. 
Interestingly, the dispersions of two sharper features in $N(q,\omega)$ can be inferred from the features noted in the corresponding spectral function shown in Fig. \ref{fig:Akw}(c). For example, the sharp low-energy mode located near the zone center can be linked to particle-hole scattering within the flat features in the spectral function near $E_F$. These correspond to the quasi-particle-like excitations with a weakly ferromagnetic character identified in the spectral function analysis of the previous section. Their bandwidth is of the same order of magnitude of the exchange coupling $J=0.25$ eV, which is somehow is reminiscent of the quasiparticle band found in 2D $t$-$J$ and Hubbard models upon hole doping~\cite{Poilblanc1993,Dagotto1994, Preuss1995}. 
Similarly, the higher energy feature in $N(q,\omega)$ can be associated with particle-hole excitations from the branch of the spectral function at high binding energies to the portion near the Fermi level. These results indicate that the substantial reorganization of the spectral function, in this case, can also be found in the dynamical charge structure factor. Later, we will show that they also appear in the spin-conserving channel of Cu $L$-edge RIXS as well. 

It is important to emphasize that several excitations observed in Fig. \ref{fig:Nqwt1t2model}(b)-(c) overlap with the ones observed in Figs. \ref{fig:Sqwt1t2model}(b)-(c), indicating that these excitations have both spin and charge character. The fact that the same excited state appears in both the spin and charge response reflects the coupling between the fractionalized spinon and holon modes and the resulting spin-polaron. This interpretation also helps explain why the reorganization of $S(q,\omega)$ and $N(q,\omega)$ is stronger when $t^\prime = \tfrac{t}{2}$, as this case has a more well defined spin-polaron. These results confirm the predictions made in Refs. \onlinecite{Eder1997,PrelovekPRB1991} but now on much larger chains with a significantly improved momentum resolution.  

\begin{figure*}[t]
    \centering
    \includegraphics[width=\textwidth]{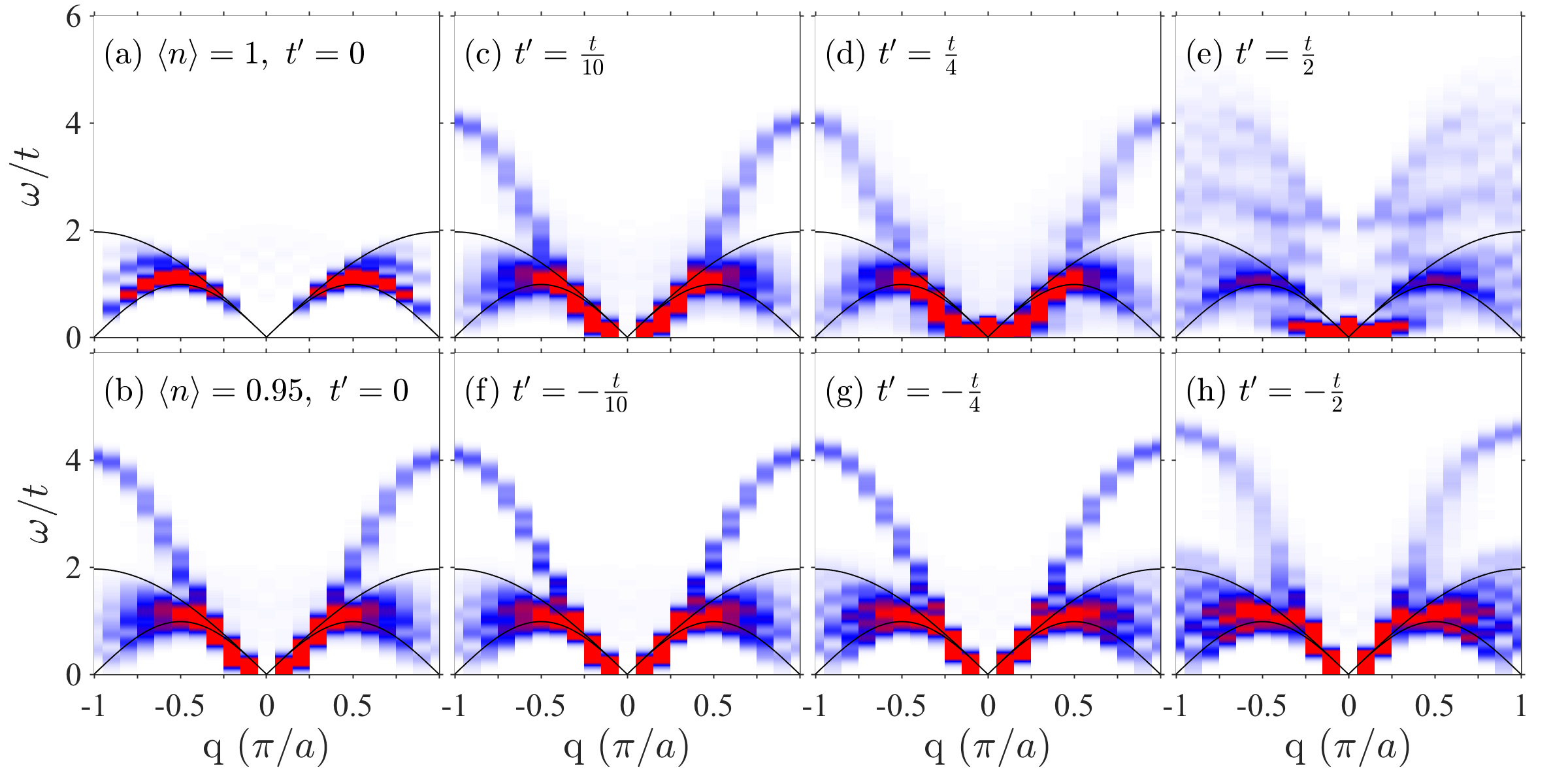}
    \caption{Exact diagonalization results for the Cu $L$-edge RIXS spectra of the $t$-$t^\prime$-$J$ model 
    in the spin-conserving ($\Delta S = 0$) channel, calculated on an $N = 20$ site chain with $t = 0.4$ eV and $J = 0.25$ eV. As with Fig. \ref{fig:rixst1t2s0}, panel (a) shows the spectra for half-filling, whereas (b)-(h) 
    show the spectra for $\langle n \rangle = 0.95$, and values of $t^\prime$ as indicated. The solid black lines indicate the boundaries of two-spinon excitations. For $t^\prime/t > 0$, the dispersion of the holon excitation appears to soften slightly until $t^\prime =\tfrac{t}{2}$ where the spectral weight completely reorganizes. For $t^\prime/t<0$, the holon exictation remains well defined and harden as the magnitude  $|t^\prime|$ increases. All panels in this figure have been plotted with the same color scale.}
    \label{fig:rixst1t2s0}
\end{figure*}

\subsection{Cu $L$-edge RIXS Spectra}\label{sec:CuLedge}
RIXS is capable of measuring collective charge and spin excitations in a single experiment \cite{AmentRMP2010}. This technique can provide a unique view of fractionalized  \cite{fsNatComm,UKUMARNJP2018, Nature.485.82} excitations in quasi-1D systems. We, therefore, computed the Cu $L$-edge RIXS response for the $t$-$t^\prime$-$J$ model for a set of $t^\prime = \{0, \tfrac{t}{10}, \tfrac{t}{4}, \tfrac{t}{2}\}$ to provide a guide for future experiments seeking to study the phenomena discussed in the previous sections. 

The RIXS response at Cu $L$-edge can be decomposed into spin-conserving (SC) ($\Delta S=0$) and non-spin-conserving (NSC) ($\Delta S = 1$) channels~\cite{PhysRevLett.103.117003,PhysRevLett.102.167401, Moretti_Sala_2011,PhysRevB.75.214414, PhysRevB.85.064422}, and each of these can be resolved by exploiting the polarization of the incoming and outgoing photons~\cite{PhysRevLett.112.147401}. We evaluated both in what follows since each gives distinct information about the system's excitations. 

\subsubsection{Cu $L$-edge RIXS in the non-spin-conserving channel}
We first consider the NSC channel. Physically, this channel is most relevant when there is a strong spin-orbit coupling in the core $2p$-orbital of the Cu site~\cite{PhysRevLett.105.157006}, and it usually dominates the Cu $L$-edge RIXS spectra. The RIXS spectra recorded in this channel also compare well with $S(q,\omega)$ at the Cu $L$-edge~\cite{PhysRevX.6.021020}. 

Fig.~\ref{fig:rixst1t2s1} shows our ED results for the Cu $L$-edge RIXS spectra in the NSC channel for the $t$-$t^\prime$-$J$ model, which were obtained using ED on an $N=20$ site chain. Panel (a) shows the spectra for half-filling ($\langle n \rangle = 1$) to apprise our readers that the spectra closely resemble $S(q,\omega)$ for the undoped chain. Here, the thin black lines indicate the boundaries of the two-spinon continuum, and our spectrum is confined within these boundaries. Panel (b) shows the RIXS spectra for $\langle n \rangle = 0.95$ and $t^\prime=0$, which compares well $S(q,\omega)$ shown in Fig.~\ref{fig:Sqwt1t2model}(a). Specifically, the sharpest feature in the spinon continuum appears to have hardened and a gapless excitation appears at $k = \pm(\tfrac{\pi}{a}-2k_\mathrm{F})$, where $2k_\mathrm{F}$ is the Fermi momentum as defined in 
Sec. \ref{sec:DCF}. 

Figs.~\ref{fig:rixst1t2s1}(c)-(e) show results $\langle n \rangle = 0.95$ and $t^\prime/t>0$, as indicated in each panel. Here, the weight of the spinon excitations near the upper boundary of the spinon continuum appears to soften for increasing values of $t^\prime$. At the same time, new spectral weight begins to appear outside of the two-spinon continuum. This additional weight is particularly pronounced at low-energies. It also becomes more prominent for $t^\prime \ge \tfrac{t}{4}$, where underlying bare Fermi surface changes from a pocket at $k=\pm\tfrac{\pi}{a}$ to two pockets centered at $k_\mathrm{min}$ [see Fig. \ref{fig:noninteractingband}(c)]. We also note that the new low-energy feature near $q = 0$ coincides with the low-energy charge excitation observed in Figs. \ref{fig:Sqwt1t2model}(c) and \ref{fig:Nqwt1t2model}(c). The fact that this feature can be resolved both in the NSC channel of RIXS and in $N(q,\omega)$ supports the interpretation that it has both spin and charge components. 

Figs.~\ref{fig:rixst1t2s1}(f)-(h) show spectra for $t^\prime/t<0$, where the spectra appear to harden slightly on increasing $|t^\prime/t|$.  (The RIXS spectra for $|t^\prime| = \tfrac{t}{4}$ shown in panels (d) and (g) compare well with the $S(q,\omega)$ results reported in Fig.~3 of Ref.~\cite{Ekaterina2019} for $|t^\prime/t| = 0.3$.) Interestingly, for intermediate values of $t^\prime$, the spectra largely resemble the $t^\prime = 0$ case. It is not until $t^\prime = -\tfrac{t}{2}$ that a new excitation appears above the spinon continuum. These results reaffirm that the formation of the spin-polaron is sensitive to the sign of $t^\prime$, and the picture emerging from Fig.~\ref{fig:rixst1t2s1} is analogous to the one obtained by examining $S(q,\omega)$ and $N(q,\omega)$ in the previous sections. 

\subsubsection{Cu $L$-edge RIXS in the spin-conserving channel}
In the case of AFM cuprates, the SC channel ($\Delta S = 0$) is dominated by charge~\cite{PhysRevX.6.021020} and double spin-flip excitations~\cite{PhysRevLett.106.157205}. Contrasting this channel with the NSC channel can, therefore, provide another avenue to probe spin and charge excitations in these systems selectively. Fig.~\ref{fig:rixst1t2s0} shows our results for the Cu $L$-edge RIXS spectra in this channel. As with the previous section, the thin black lines indicate the boundaries of the two-spinon continuum. 

Fig. \ref{fig:rixst1t2s0}(a) shows results obtained at half-filling to remind the reader of what occurs in the undoped case. Here, the spectrum is dominated by double spin-flip excitations, where the spectral weight is mostly confined to the two-spinon continuum but with a node in the intensity near the zone boundaries $q=\pm \tfrac{\pi}{a}$~\cite{PhysRevLett.106.157205, PhysRevB.83.245133, PhysRevB.85.064422, UKumar2019}. Our results are in agreement with earlier studies~\cite{PhysRevB.85.064423, UKUMARNJP2018}; however, we also observe a feature with a very weak intensity centered near $(q,\omega) = (0,t)$, which resembles four-spinon excitations uncovered previously at the oxygen $K$-edge~\cite{fsNatComm} but with diminished intensity. Fig. \ref{fig:rixst1t2s0}(a) thus confirms that these excitations can also be resolved at the Cu $L$-edge, albeit with an overall weaker intensity due to the shorter lifetime of the Cu $2p$ core-hole~\cite{fsNatComm}. 

Fig.~\ref{fig:rixst1t2s0}(b) shows the SC RIXS response for the doped case $\langle n \rangle = 0.95$ with $t^\prime = 0$.  Compared to the undoped case, one sees an additional dispersing feature with a bandwidth of $4t$. A similar excitation was predicted at the oxygen $K$-edge~\cite{UKUMARNJP2018} and was attributed to a holon excitation. This interpretation is supported by the fact that the dispersion of this excitation agrees well with that expected for a holon $\omega(q) = 2t[1-\cos(q a)]$, and with $N(q,\omega)$ computed for a dilute spinless chain [see Fig.~\ref{fig:noninteractingband}(d)]. 

Figs. \ref{fig:rixst1t2s0}(c)-(e) show spectra for $\langle n\rangle = 0.95$ and $t^\prime/t>0$ as indicated, while panels (f)-(h) show the cases with $t^\prime/t<0$. 
For $|t^\prime| < \tfrac{t^\prime}{4}$, the spectra look similar to the $t^\prime = 0$ case, but with a slight softening (hardening) of the holon excitation for $t^\prime > 0$ ($t^\prime < 0$).  However, for $t^\prime=\tfrac{t}{2}$ [panel (e)], the charge excitations break up into an two distinct sets of excitations with additional incoherent weight at high energy, similar to what was observed for the dynamical charge structure factor shown in Fig. \ref{fig:Nqwt1t2model}(c). We, therefore, associate them with the same particle-hole scattering processes within the spectral function shown in Fig. \ref{fig:Akw}(c). 

\section{Conclusions}\label{sec:conclusions}
We have studied the effects of nnn hopping on the evolution of the low-energy charge and spin dynamics of quasi-1D AFM cuprate spin chains within the doped $t$-$t^\prime$-$J$ model. Specifically, we presented DMRG results for the single-particle spectral function and the dynamical spin and charge structure factors. We found evidence that $t^\prime$ couples the fractionalized spinon and holon excitations. This coupling can lead to a spin-polaron state, where a ferromagnetic spin polarization cloud dress the doped electron. In this respect, our results are consistent with an earlier ED study that was carried on $N = 16$ site chains \cite{Eder1997} but with higher momentum resolution and covering a wider parameter regime. As such, we can obtain a complete picture of the breakdown of fractionalization with the inclusion of nnn hopping. To the best of our knowledge, the charge dynamics of the doped 1D $t$-$t^\prime$-$J$ model has not been widely explored in the existing literature. 

Our results provide details predictions for the dynamical response functions of doped cuprate spin chains with nnn hopping, which can be probed by spectroscopies such as INS or ARPES. We also provided predictions for the Cu $L$-edge RIXS spectra. RIXS has emerged as a novel spectroscopic method for probing both charge and spin excitations in quantum magnets within a single experiment. It is, therefore, an ideal probe for exploring fractionalization in 1D~\cite{Nature.485.82,fsNatComm,UKUMARNJP2018}. We evaluated both the NSC and SC channels at this edge and identified excitations related to fractionalized spinons and holons, as well as the spin-polaron as a function of $t^\prime$. Based on this, we propose that Cu $L$-edge RIXS measurements can be used to identify the presence of spin-polarons in various spin-chain systems with appropriate values of $t^\prime$. 1D chain cuprates compound such as SrCuO$_2$~\cite{KarmakarCGD2014} and Sr$_2$CuO$_3$~\cite{KarmakarCGD2015} can be doped; however, the 
magnitude of $t^\prime$ in the corner-shared system Sr$_2$CuO$_3$ should be small and so the zig-zag SrCuO$_2$ may be a better candidate. 
In this context, the edge-shared cuprates may be another preferable system as the Cu-O-O-Cu hybridization pathways should result in larger effective $t^\prime$ than the corner-shared cuprates, which tend to be dominated by nn Cu-O-Cu hopping processes. Another possibility may be to engineer 1D iridates with long-range hopping, which has recently been shown to be possible in artificial heterostructures~\cite{doi:10.1002/adma.201603798}.

\begin{acknowledgments}
U.~K. acknowledges partial support from US DOE NNSA under contract No. 89233218CNA000001 through the LDRD Program. G.~P. acknowledges funding support from 2018 Augusta University CURS summer scholars program. A.~N. acknowledges support from the Canada First Research Excellence Fund. S.~J. acknowledges support from the National Science Foundation under Grant No. DMR-1842056. T.~D. acknowledges funding support from Augusta University Scholarly Activity Award and Sun Yat-Sen University Grant No. OEMT-2017-KF-06. This work used computational resources supported by the University of Tennessee and Oak Ridge National Laboratory Joint Institute for Computational Sciences, and additional computational resources and services provided by Advanced Research Computing at the University of British Columbia.
\end{acknowledgments}

\bibliographystyle{apsrev4-1}
\bibliography{1dtjrixs}
\end{document}